\pdfoutput=1

\documentclass[11pt,a4paper]{article}
\usepackage{jcappub}

\usepackage{epsfig,psfrag,graphics,verbatim}
\usepackage{graphicx}
\usepackage{epstopdf}
\usepackage{dcolumn}
\usepackage{bm}
\usepackage{slashed}
\usepackage{color}
\usepackage{amssymb}

\begin{document} 
\hfill{{\small TUM-HEP 867/12}}

\title{Taming astrophysical bias in direct dark matter searches}

\author[a]{Miguel Pato,}
\author[b]{Louis E. Strigari,}
\author[c]{Roberto Trotta}
\author[d]{and Gianfranco Bertone}

\affiliation[a]{Physik-Department T30d, Technische Universit\"at M\"unchen, James-Franck-Stra\ss{}e, 85748 Garching, Germany}
\affiliation[b]{Kavli Institute for Particle Astrophysics and Cosmology, Stanford University, Stanford, CA 94305 USA}
\affiliation[c]{Astrophysics Group \& Imperial Centre for Inference and Cosmology, Imperial College London, Blackett Laboratory, Prince Consort Road, London SW7 2AZ, UK} 
\affiliation[d]{GRAPPA Institute, University of Amsterdam, Science Park 904, 1090 GL Amsterdam, Netherlands}

\emailAdd{miguel.pato@tum.de}
\emailAdd{strigari@stanford.edu}
\emailAdd{r.trotta@imperial.ac.uk}
\emailAdd{gf.bertone@gmail.com}

\date{\today}

\abstract{
We explore systematic biases in the identification of dark matter in future direct detection experiments and compare the reconstructed dark matter properties when assuming a self-consistent dark matter distribution function and the standard Maxwellian velocity distribution.  We find that the systematic bias on the dark matter mass and cross-section determination arising from wrong assumptions for its distribution function is of order $\sim 1 \sigma$.  A much larger systematic bias can arise if wrong assumptions are made on the underlying Milky Way mass model. However, in both cases the bias is substantially mitigated by marginalizing over galactic model parameters. We additionally show that the velocity distribution can be reconstructed in an unbiased manner for typical dark matter parameters. Our results highlight both the robustness of the dark matter mass and cross-section determination using the standard Maxwellian velocity distribution and the importance of accounting for astrophysical uncertainties in a statistically consistent fashion.
}

\maketitle

\section{Introduction}

\par Determining the properties of weakly interacting massive particles (WIMPs) from the variety of available experimental techniques presents an exciting experimental and theoretical challenge \cite{BertoneBook,Jungman,Bergstrom,Munoz:2003gx,Bertone:2004pz}. Direct, indirect, and collider dark matter searches each come with their own unique set of benefits from a theoretical perspective, as well as systematic issues from an experimental perspective. A solution to the dark matter problem certainly requires convergence from each of these major techniques (see e.g.~Refs.~\cite{Bertone:2007xj,Moulin:2007ij,Mambrini:2009zz,Bertone:2010rv,Pato:2010zk,Bergstrom:2010gh,Bertone:2011pq,Nguyen:2012rx,Mambrini:2012ue}). 

\par This paper focuses on understanding how accurately WIMP properties -- in particular the mass and elastic spin-independent (SI) scattering cross-section -- can be determined from direct dark matter detection experiments given systematic uncertainties coming from our imperfect knowledge of the dark matter astrophysical distribution. Several recent studies have tackled this issue in one form or another: Strigari \& Trotta \cite{Strigari:2009zb} introduced a Bayesian method to determine WIMP properties by using kinematic data in the Milky Way and marginalizing over galactic halo model parameters; this method was then improved in Ref.~\cite{Pato:2010zk} to consider complementarity between different experimental targets for a more general parameterization of the velocity distribution \cite{Lisanti:2010qx}. Using mock data from future direct detection experiments, Peter \cite{Peter:2011eu} (see also \cite{Kavanagh:2012nr}) identified biases in the WIMP mass determination for different phenomenological velocity distribution models, determining both the circular and escape velocities from the event rate distribution, while \cite{Strege:2012kv} investigated systematic biases in the determination of WIMP properties arising from unavoidable statistical fluctuations in the energy distribution.

\par Here we extend the above studies and consider the bias in the reconstructed WIMP properties, improving on previous work in two specific ways. First, we study bias for several theoretically well-motivated models of the smooth WIMP velocity distribution. In particular, we apply the above techniques to theoretical models that establish a direct correspondence between the velocity distribution and the dark matter density profile in a self-consistent manner. While there have been several works that have explored phenomenologically how these models impact the observed event rate \cite{Ullio:2000bf,Evans:2005tn,Lisanti:2010qx,Catena:2011kv,Arina:2011si}, we examine for the first time the bias in the reconstruction of WIMP properties for these models (see also \cite{Bhattacharjee:2012xm}). Second, we study the bias in the WIMP mass and cross-section determination while simultaneously fitting kinematic data that probe the dark matter distribution of the Galaxy.  This approach accounts for the effect of uncertainties in galactic model parameters in a statistically robust fashion, and we argue that this or a similar technique will be necessary in the case of a confirmed WIMP signal. 

\par We envision that the approach undertaken in this paper extends beyond the ``discovery" phase of dark matter detection, anticipating a period when robust statistical procedures will be required to extract WIMP properties in an accurate way from direct detection data. Our approach is complementary to recent ones \cite{Fox:2010bz,Frandsen:2011gi,Gondolo:2012rs} that have been developed to elucidate whether the events seen by multiple experiments \cite{DAMA2008,DAMA2010,cogentannualmod,Angloher:2011uu} can be consistent with a WIMP signal. These approaches encapsulate the impact of the astrophysical dark matter distribution in one single galactic halo model parameter that determines the event rate distribution and are of particular interest if several experiments are indeed seeing WIMP signals. 

\par The main results of this work quantify the bias that is incurred for different models of the velocity distribution, both with and without including uncertainties in galactic model parameters. While we show that the WIMP mass and cross-section are systematically unbiased when assuming the true underlying velocity distribution (as one would expect), standard reconstruction methods do not recover the true event rate distribution when assuming an incorrect model of the velocity distribution. Nevertheless, we find that it is straightforward to overcome this bias by marginalizing over galactic model parameters, and that the use of the standard Maxwellian velocity distribution leads to a robust determination of the WIMP properties from direct detection experiments.

\par The outline of the paper is as follows. In Section \ref{secMWDD} we present the modeling of the different Milky Way components, while Section \ref{secDD} details the calculation of the WIMP event rates. Section \ref{secMeth} is devoted to outlining our methodology. Our main results are shown and discussed in Section \ref{secRes}, before concluding in Section \ref{secConc}.

\section{Milky Way modelling}\label{secMWDD}

\par In this section we motivate the models for the Milky Way that we consider, and relate these to the corresponding WIMP velocity distributions.

\subsection{Standard halo model}

\par For our main analysis we consider spherically symmetric and isotropic dynamical models. Though it is straightforward to consider more complicated models, the above assumptions are appropriate for the interpretation of the primary results of this paper. Let us define $\vec v$ as the WIMP velocity in the Earth rest frame, $\vec{v}_\oplus$ as the Earth velocity in the galactic rest frame, and  $\vec{w}\equiv\vec{v}+\vec{v}_\oplus$ as the WIMP velocity in the galactic rest frame. The Maxwellian distribution truncated at the escape velocity is 
\begin{equation}\label{Maxwell}
f({\vec{w}})\propto \left[ \exp\left(\frac{v_{esc}^2-w^2}{v_0^2}\right) -1 \right] \quad,
\end{equation}
for $w<v_{\text{esc}}$ and 0 otherwise, with given values for $v_0$ and $v_{\text{esc}}$. This velocity distribution, or its variations, are often referred to as the standard halo model (SHM). We take $\vec{v}_\oplus \simeq v_0$ and define $v_0$ as the local circular velocity, which is derived from the total galactic potential, $\phi_{\text{tot}}$, as 
\begin{equation}
v_0^2 = R_\odot \left. \frac{d \phi_{\text{tot}}}{dr} \right|_{r=R_\odot} \quad ,
\end{equation}
where $R_\odot$ is the galactocentric distance. Similarly, the escape velocity is derived from the total potential as
\begin{equation}
v_{\text{esc}}^2 = 2\left|\phi_{\text{tot}}(R_\odot)\right|  \quad.
\end{equation}
To determine $\phi_{\text{tot}}$ we consider a Milky Way model split into three separate components, the bulge ($b$), the disk ($d$) and the dark matter halo ($dm$), and sum the contribution from each component, 
\begin{equation}\label{gravpot}
\phi_{\text{tot}} = \phi_{b} + \phi_d + \phi_{dm} \quad. 
\end{equation}
For the bulge we take the spherical potential $\phi_b=-G M_b/(r+c_0)$ and for the disk we take the spherical approximation $\phi_d=-G M_d[1-\exp(-r/b_d)]/r$, where $M_b=1.5\times 10^{10}\textrm{ M}_\odot$ is the fixed bulge mass, $c_0=0.6$ kpc is the fixed bulge scale radius, $M_d$ is the disk mass and $b_d$ is the disk scale radius (both treated as a free parameters, as explained later), see Strigari \& Trotta \cite{Strigari:2009zb} for references and further details. We specify the dark matter component by its mass density that is taken to follow a generalized Navarro-Frenk-White (NFW) profile:
\begin{equation}\label{NFW}
\rho_{dm}=\frac{\rho_s}{(r/r_s)^a (1+(r/r_s)^b)^{(c-a)/b}} \quad ,
\end{equation}
where $a$, $b$ and $c$ are profile indices, $\rho_s$ ($r_s$) is the scale density (radius) and $\rho_0\equiv \rho_{dm}(R_\odot)$ is the local dark matter density.

\par With the potential specified, the radial velocity dispersion for a given stellar population is determined from the spherical and isotropic Jeans equation,
\begin{equation}
\frac{\partial \left(\rho_\star \sigma_r^2\right)}{\partial r} = -G\rho_\star \frac{\partial \phi_{\text{tot}}}{dr} \quad, 
\label{jeans} 
\end{equation}
where $\rho_\star$ is the density distribution for the stellar population. As motivated in more detail below, we will assume that the population of stars is isotropic in velocity space. In the limit of an isotropic velocity distribution, the radial velocity dispersion $\sigma_r$ from Eq.~\eqref{jeans} is identical to the observed line-of-sight velocity dispersion. 

\par In this model, in order to evaluate Eq.~\eqref{Maxwell}, there are thus two parameters that must be determined from the kinematic data on the Galaxy, namely $v_0$ and $v_{\text{esc}}$. These are functions of the entire set of parameters that determine the total local potential, i.e.~$\{\rho_s,r_s,a,b,c,b_d,M_d,R_\odot\}$. These parameters are fit to the kinematic data described below. Once this is done, the WIMP event rate distribution is derived from the velocity 
distribution function in Eq.~\eqref{Maxwell}. 

\par The modeling described above is very similar to that used in Strigari \& Trotta \cite{Strigari:2009zb}, who connected the WIMP event rate distribution to the kinematics of halo stars and other measurements of galactic model parameters. The only difference between what we have outlined above and the analysis by Strigari \& Trotta is that here we set the anisotropy parameter $\beta$ to zero. However, in this paper we extend the former formalism to study a self-consistent model that connects the dark matter density distribution with the corresponding velocity distribution. We now proceed to describe this model.

\subsection{Self-consistent distribution function}\label{sub:self}

\begin{figure*}
\begin{center}
\begin{tabular}{c}
\includegraphics[width=0.5\textwidth]{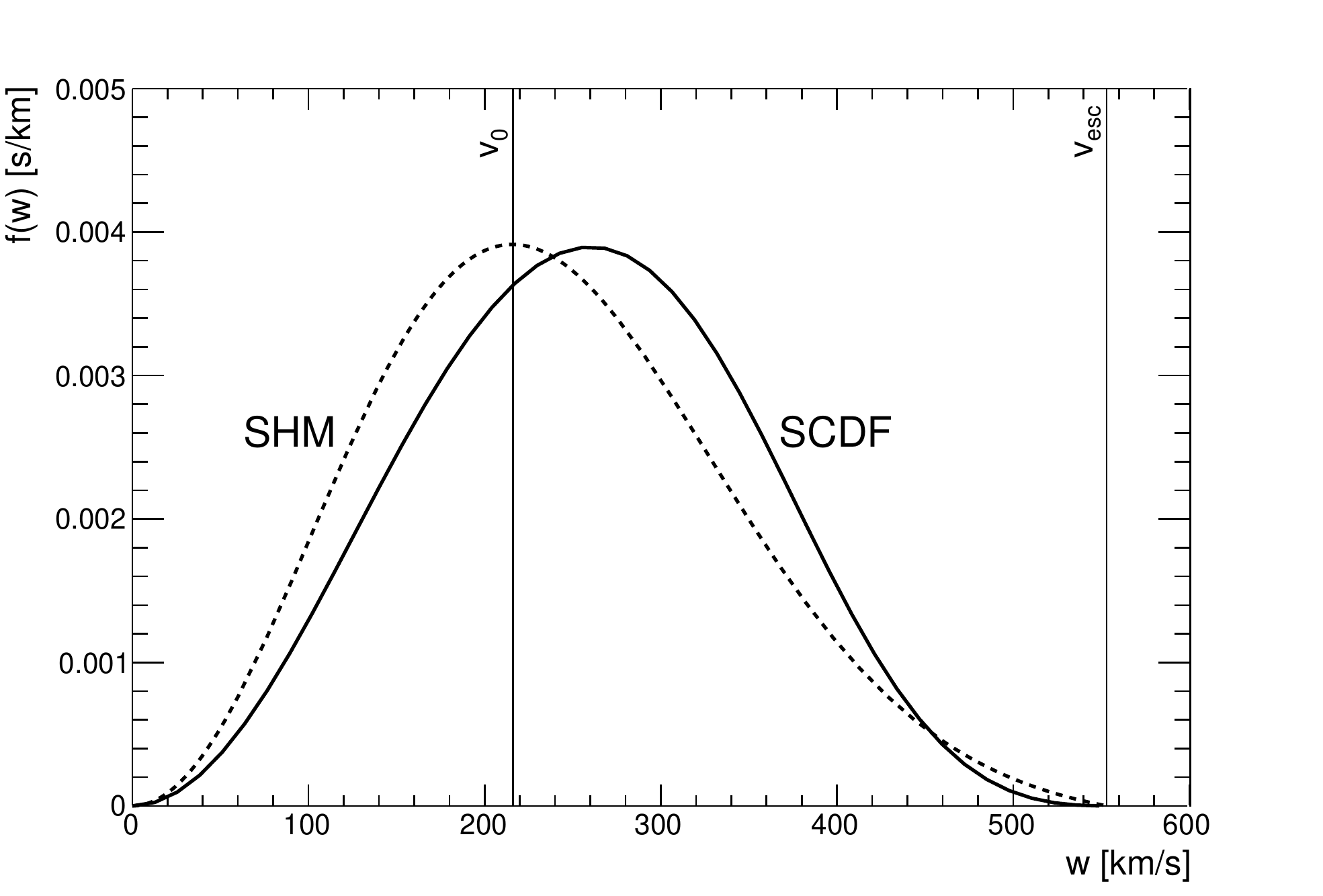}
\includegraphics[width=0.5\textwidth]{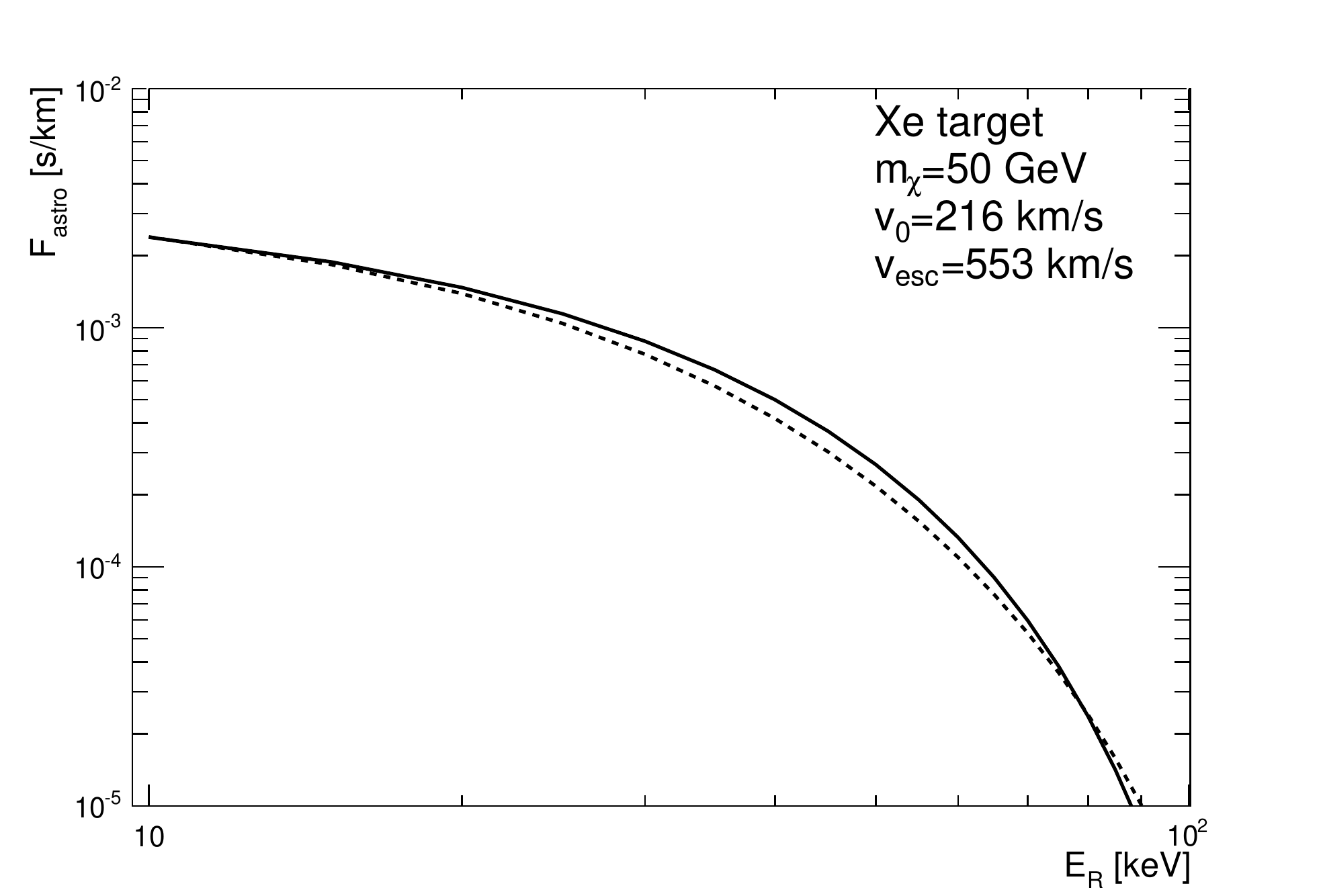}
\end{tabular}
\end{center}
\caption{The WIMP velocity distribution (left) and the mean inverse velocity (right) for the case of the standard halo model (dashed lines) and the self-consistent distribution function (solid lines) with the fiducial values in Table \ref{tab:models}. The mean inverse velocity $F_{\textrm{astro}}$ in the right frame corresponds to the velocity integral in Eq.~\eqref{rate} for a 50 GeV WIMP and a xenon target.
}
\label{fig:comparison}
\end{figure*}

\par Though the above procedure for determining the velocity distribution from $v_0$ and $v_{\text{esc}}$ is well defined, it is not necessarily dynamically self-consistent, in that the velocity distribution in Eq.~\eqref{Maxwell} does not follow from the bulge, disk, and dark matter potentials. It is thus important to compare the SHM to a self-consistent model, in which the distribution function is generated from an one-to-one mapping to the dark matter density profile. To perform this comparison, we consider the Eddington formula
\begin{equation}\label{Eddington}
f(\epsilon) = \frac{1}{\sqrt{8}\pi^2}\int_0^\epsilon{\frac{\partial^2 \rho_{dm}}{\partial \Psi_{\text{tot}}^2} \frac{d\Psi_{\text{tot}}}{\sqrt{\epsilon-\Psi_{\text{tot}}}}  } \quad,
\end{equation}
which is valid under the assumption of spherical symmetry and isotropy. Here, $\epsilon$ is the relative specific energy of the WIMP and reads
\begin{equation}
\epsilon\equiv -E(R_\odot)+\phi_{\text{tot}}(\infty)=-w^2/2+\Psi_{\text{tot}}(R_\odot)
\end{equation}
with $E(R_\odot)=w^2/2+\phi_{\text{tot}}(R_\odot)$ and $\Psi_{\text{tot}}(R_\odot)=-\phi_{\text{tot}}(R_\odot)+\phi_{\text{tot}}(\infty)$. For an assumed parameter set that determines the potential, the self-consistent distribution function (SCDF) derived from Eq.~\eqref{Eddington} differs from the SHM distribution. The crux of our work lies precisely in the comparison between the velocity distributions obtained from Eqs.~\eqref{Eddington} and \eqref{Maxwell} -- both are plotted in Fig.~\ref{fig:comparison} (left). It is interesting that, although the distributions are clearly different, they are both featureless and share a similar overall shape -- this will have important consequences in our findings as discussed in Section \ref{secRes}. Let us notice at this point that the underlying astrophysical model -- defined by the parameter set $\{\rho_s,r_s,a,b,c,b_d,M_d,R_\odot\}$ -- is independent of the assumptions on the velocity distribution. These assumptions, however, affect crucially direct detection rates, as illustrated later on.

\par Though deriving the velocity distribution self-consistently from the density profile is in principle more theoretically appealing than simply using the SHM, a few caveats are in order. First, it appears that the velocity distribution from cosmological simulations of dark matter halos differs significantly from the correspondence from Eddington theory \cite{Vogelsberger:2008qb,Kuhlen:2009vh,Read:2009iv,Mao:2012hf,Kuhlen:2012fz}. Hence, this model itself may still be an oversimplification if the phase space properties of simulated halos are a reflection of the phase space properties of the Milky Way. Nonetheless, the comparison between the SHM and the SCDF has important phenomenological implications, as we show in detail below.

\section{Direct Detection}\label{secDD}

\par Our reconstruction of dark matter properties from simulated direct detection data follows standard analysis procedures; hence we only briefly outline the relevant points in this Section. The crucial galactic ingredient for direct detection is the dark matter phase space distribution, more precisely the local dark matter density $\rho_0$ and local velocity distribution $f(\vec{w})$. In fact, the scattering rate induced by WIMPs of mass $m_\chi$ and SI WIMP-proton cross-section $\sigma_p^{\text{SI}}$ on target nuclei $N(A,Z)$ is given by
\begin{equation}\label{rate}
\frac{dR}{dE_R} = \frac{\rho_0 \sigma_p^{\text{SI}}}{2 m_\chi \mu_p^2} A^2 F_{\text{SI}}^2(E_R) \int{ \frac{d^3\vec{v}}{v} \, f(\vec{w}) \Theta(v-v_{\text{min}})  } \quad ,
\end{equation}
where $\mu_p$ is the WIMP-proton reduced mass, $v_{\text{min}}=(m_N E_R/(2\mu_N^2))^{1/2}$ is the minimum velocity of the WIMP in the Earth frame to produce a nuclear recoil of energy $E_R$ and $F_{\text{SI}}$ is the spin-independent form factor (see \cite{Pato:2010zk} for further details). Eq.~\eqref{rate} assumes elastic SI-nucleon scattering only; although other types of scattering have been extensively discussed in the literature, we focus on the elastic SI case for concreteness.

\section{Methodology}\label{secMeth}

\begin{table*}
\caption{\label{tab:models}Summary of the parameters used in the analysis. We have assumed uniform priors within the given ranges. The letter in the last column refers to the setups in which these parameters are used: f (fixed), p (pheno), and F (full). The setups are defined in Section \ref{secMeth}.}
\fontsize{9}{9}\selectfont
\begin{tabular}{@{}l|l|c|c|c}
\hline
\hline
Symbol & Parameter & Fiducial values & Prior range & Model used \\
\hline 
\hline 
$\log_{10} \left[m_\chi/\textrm{GeV}\right]$ &		WIMP mass	&	$\log_{10}(25,50,250)$ &	$0.1-3.0$ &		f,p,F \\
$\log_{10} \left[\sigma_p^{\text{SI}}/\textrm{pb}\right]$ & 	spin-independent cross-section &	$-9, -8.5$ &			$-11-(-7)$ &		f,p,F\\
\hline
$\rho_0 \left[\textrm{GeV}/\textrm{cm}^{3}\right]$ &local dark matter density&	$0.29$ &			$0.001 - 0.79$ & 	p \\
$v_0$ [km/s] & 		circular velocity&				$216$ & 			$66 - 366$ & 		p \\
$v_{\textrm{\text{esc}}}$ [km/s] & 		escape velocity&	 			$553$ &				$388 - 718$ & 		p \\
\hline
$\log_{10} \left[\rho_s/(\textrm{M}_\odot/\textrm{kpc}^{3})\right]$& 		scale density & 	$6.8$ & 	$5-8$ & 		F \\
$r_s$ [kpc] & 					scale radius &	$20$ &				$1-60$ & 		F \\
a &							inner slope & $1$ & 				$0-2$ & 		F 	\\
b & 							transition slope &	$1$ & 				$0-2$ & 		F \\
c & 							outer slope	&$3$ & 				$2-5$ &		F 		\\
$b_d$ [kpc] & 					disk scale radius & 		$4$ & 				$2-6$ &			F \\
$M_d$ [$10^{10}$ M$_\odot$] &   	disk mass		&	$7$ &				$2-8$ &	F \\
$R_\odot$ [kpc] & 				galactocentric distance		&		$8.33$ &				$7-10$ &		p,F 	\\
\hline 
\end{tabular}
\end{table*}

\subsection{Likelihood and data} 

\par We adopt a Bayesian methodology and employ the MultiNest code \cite{Feroz:2007kg,Feroz:2008xx,Trotta:2008bp} to derive samples from the posterior probability distribution function (pdf) $p(\Theta|d)$ for the parameter set $\Theta$ given the data $d$ using Bayes' theorem:
\begin{equation}
p(\Theta|d)\propto \mathcal{L}(\Theta) p(\Theta) 
\end{equation}
with $p(\Theta)$ the prior pdf (which is taken to be uniform within a certain range) and $\mathcal{L}(\Theta)$ the likelihood function. In the case at hand, $\Theta$ includes both the astrophysical parameters discussed in Section \ref{secMWDD} and the WIMP properties $m_\chi$ and $\sigma_p^{\text{SI}}$ -- see Table \ref{tab:models}, where the adopted prior ranges and fiducial values for each parameter are shown. 

\par We consider three separate setups, which differ in the number of parameters that are included in $\Theta$, the data sets used, and the models for the velocity distribution used:

\bigskip 

$\bullet$ The first and simplest setup, which we label the {\em fixed} model (f), varies only the WIMP mass and cross-section, hence
\begin{equation}
 \Theta = \{m_\chi, \sigma_p^{\text{SI}}\}\quad .
\end{equation}
The galactic model parameters are fixed to their given fiducial values, whose motivation is discussed in more detail below.

\bigskip

$\bullet$ The second setup, which we refer to as the phenomenological (i.e.~{\em pheno}, p) model, varies the local dark matter density, local circular velocity, escape velocity, and galactocentric distance in addition to the WIMP mass and cross-section, so that
\begin{equation}
 \Theta = \{m_\chi, \sigma_p^{\text{SI}}, \rho_0, v_0, v_{\text{esc}},R_\odot\}\quad .
\end{equation}
In this setup, we use the measurements of the local dark matter density \cite{Catena:2009mf,Salucci:2010qr,Pato:2010yq}, circular velocity \cite{Xue:2008se,Sofue:2008wu,Reid:2009nj,Bovy:2009dr,McMillan:2009yr,Pato:2010zk}, escape velocity \cite{Smith:2006ym}, and galactocentric distance \cite{Gillessen:2008qv}, and adopt a Gaussian likelihood with the following means and standard deviations:
\begin{eqnarray}
\rho_0 &=& 0.29 \pm 0.1\textrm{ GeV/cm}^3 \nonumber \\ 
v_0 &=& 216 \pm 30 \textrm{ km/s} \nonumber \\
v_{\text{esc}} &=& 544\pm 33\textrm{ km/s} \nonumber \\
R_\odot &=& 8.33 \pm 0.35 \textrm{ kpc}  \quad ,
\end{eqnarray}
where the central values for $\rho_0$ and $v_0$ are the ones obtained in our Milky Way fiducial model, see below and Table \ref{tab:models}.

\bigskip 

$\bullet$ The third setup ({\em full}, F) allows the largest flexibility in the galactic model parameters, as defined in Section \ref{secMWDD}, by treating all of them as free parameters along with the WIMP mass and cross-section. In this setup, the parameter set is  
\begin{equation}
\Theta=\{m_\chi, \sigma_p^{\text{SI}},\rho_s,r_s,a,b,c,b_d,M_d,R_\odot\}\quad. 
\end{equation}
In addition to the above constraints on $v_{\text{esc}}$ and $R_{\odot}$, we add to the likelihood the constraint on the velocity dispersion profile of the Milky Way halo, utilizing the radial velocity dispersion of 2400 blue horizontal branch (BHB) halo stars as measured by the Sloan Digital Sky Survey (SDSS) \cite{Xue:2008se}. In the isotropic Jeans equation, we take the stellar density profile of the tracer population to scale as $\rho_\star\propto r^{-3.5}$ for $r>10$ kpc and $\rho_\star\propto\textrm{const}$ otherwise. To implement Eq.~\eqref{jeans}, we use the measured velocity dispersion by Xue et al \cite{Xue:2008se} into 9 radial bins, and assume an independent Gaussian likelihood for each bin. 

\bigskip 

\par Note that, in the {\em full} setup, $\rho_0$, $v_0$, and $v_\textrm{\text{esc}}$ are derived parameters, and do not appear as an input into either the model for the smooth density distribution or in the velocity distribution. This is in contrast to the {\it pheno} setup, in which these parameters directly appear. In both the {\em full} and the {\em pheno} setups, $\rho_0$, $v_0$, and $v_\textrm{\text{esc}}$ are constrained from the observations.  

\par The joint likelihood function $\mathcal{L}(\Theta)$ is the product of a term associated with galactic parameters and a term coming from direct detection data:
\begin{equation}
\mathcal{L}(\Theta) = \mathcal{L}_\text{astro}(\Theta) \times \mathcal{L}_\text{DD}(\Theta)\quad.
\end{equation}
This is the likelihood function used in all the scans presented in Section \ref{secRes} in order to infer the posterior in the parameter space $\Theta$. The likelihood for the galactic parameters, $\mathcal{L}_\text{astro}$, is a multinormal Gaussian as described above. The direct detection likelihood, $\mathcal{L}_\text{DD}$, is computed as follows. First, we determine the fiducial values of the astrophysical parameters by performing a scan of the full model, using for the astrophysics likelihood the constraints for $R_{\odot}$, $v_{\text{esc}}$ as well as the kinematics of SDSS halo stars discussed above.  The mean of the posterior distribution over the astrophysical parameters is then used to approximately fix the fiducial values for each astrophysical parameter given in Table \ref{tab:models}. This defines our assumed true galactic model (i.e., the fiducial model), making sure that it is close to what current astrophysical data require. We then generate mock direct detection counts through Eqs.~\eqref{rate} and \eqref{Eddington} from the fiducial model, assuming a xenon, germanium or neon experiment with 2.00, 2.16 and 150 ton$\times$yr of effective exposures respectively (see Refs.~\cite{Pato:2010zk,Hime:2011ms}) and several benchmark WIMP parameters. The energy resolution for the xenon and germanium cases is as in \cite{Pato:2010zk}, while for neon we take the energy-dependent resolution $\sigma(E_R)=1.0\textrm{ keV}\sqrt{E/\textrm{keV}}$. The counts are distributed into 10 linearly-spaced bins in the range $E_R=10-100$ keV for xenon and germanium and $E_R=20-100$ keV for neon, and  $\mathcal{L}_\text{DD}$ is a standard binned likelihood. Since we are interested in studying systematic biases arising from assumptions made about the dark matter distribution function, we generate mock counts without Poisson noise (rounding mock counts to the nearest integer, an approximation that is unlikely to have an impact on our results given the large number of counts expected in most energy bins). This procedure eliminates statistical fluctuations from our realised data, so that -- in absence of systematic errors -- our maximum a posteriori estimate of the parameters value ought to coincide with the true input value.

\subsection{Systematic bias from astrophysical assumptions}

\par With the ingredients outlined in the previous paragraphs we aim to quantify the systematic bias caused by wrong assumptions for the galactic halo model.  We thus run several scans attempting a reconstruction of the dark matter mass and cross-section from the above-described direct detection data, which were simulated from an astrophysical model obeying the Eddington formula \eqref{Eddington}, but reconstructing the WIMP properties adopting the truncated Maxwellian distribution \eqref{Maxwell}. If the Eddington formula \eqref{Eddington} and the truncated Maxwellian distribution \eqref{Maxwell} yield similar counts, or if astrophysical and statistical uncertainties are large, the posterior distribution should be well-centered about the true WIMP properties and the corresponding systematic bias is negligible. However, if that is not the case, then making the wrong assumption about the velocity distribution, i.e.~adopting Eq.~\eqref{Maxwell}, will systematically bias the inference from direct detection data. 

\par In order to quantify this effect we define our ``bias factor'' $\alpha$ as 
\begin{equation}\label{bias}
\alpha = \sqrt{\left(\frac{d_1}{\sqrt{\lambda_1}}\right)^2+\left(\frac{d_2}{\sqrt{\lambda_2}}\right)^2} 
\end{equation}
with $d_{1,2}=\left((x_t,y_t)-(\bar{x},\bar{y})\right)\cdot \vec{\lambda}_{1,2}/|\vec{\lambda}_{1,2}|$, where $\lambda_1$ and $\lambda_2$ ($\vec{\lambda}_1$ and $\vec{\lambda}_2$) are the eigenvalues (eigenvectors) of the covariance matrix in the parameter space $(x,y)\equiv(\log_{10}\left[m_\chi/\textrm{GeV}\right],\log_{10}\left[\sigma_p^{\text{SI}}/\textrm{pb}\right])$, $(x_t,y_t)$ indicates the true WIMP properties and $(\bar{x},\bar{y})$ the posterior mean. The bias factor $\alpha$ simply measures the distance between the true and mean points in units of standard deviations along the eigenvectors of the covariance matrix. In the following we shall describe a particular scan as biased if $\alpha>1$.

\section{Results \& Discussion}\label{secRes}

\par
\begin{figure}
\begin{center}
\begin{tabular}{c}
\includegraphics[width=0.6\textwidth]{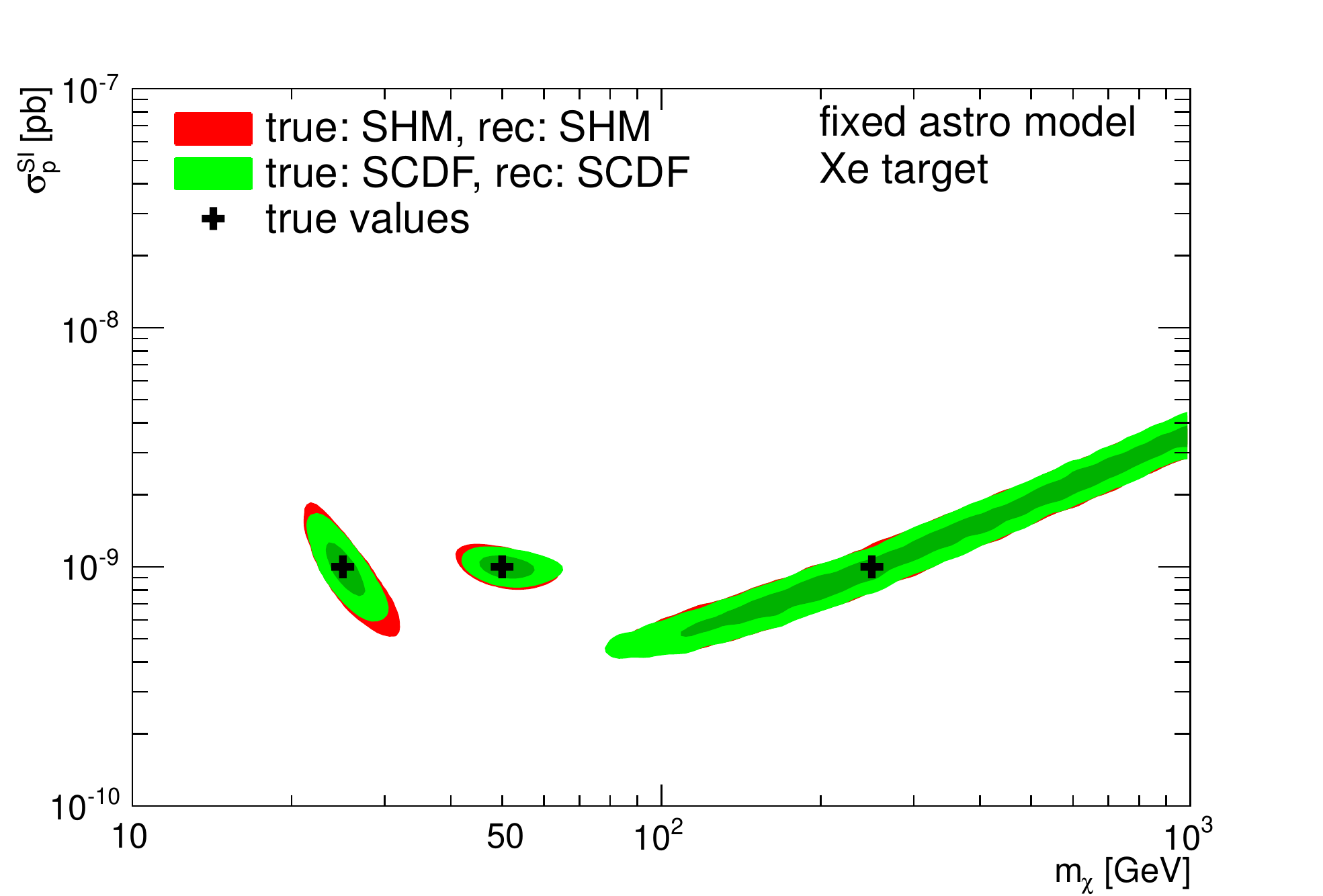}
\end{tabular}
\end{center}
\caption{Unbiased reconstruction of WIMP properties in a xenon ton-scale instrument for the fixed astrophysical setup, when the distribution function used for the reconstruction (``rec'') matches the one used to generate the data (``true''). Inner and outer contours correspond to the joint 68\% and 95\% posterior probability contours, respectively. The red (dark) contours assume the standard halo model both in generating the mock data and in the reconstruction, while the green (light) contours assume the self-consistent distribution function.
}
\label{fig:true}
\end{figure}

 Firstly, we investigate the statistical accuracy with which the WIMP mass and cross-section can be determined in the absence of systematic bias, i.e., if we assume that we know the correct velocity distribution model. The reconstruction is shown in Fig.~\ref{fig:true}, for three benchmark values of the WIMP mass. In each case, we simulate counts assuming either the SHM or the SCDF, and we reconstruct mass and cross-section assuming the correct distribution model {\em and} correct, fixed parameters for it. This is of course an overly optimistic scenario, but it delineates the maximum statistical accuracy that can be achieved in the absence of astrophysical uncertainties. As shown by Fig.~\ref{fig:true}, the determination of the WIMP mass and cross-section is roughly equivalent for both the SHM and SCDF in this optimistic scenario. For each fiducial set of parameters, there is no significant systematic bias, i.e.~the fiducial parameters are well-determined at the one-sigma level. 

\par
\begin{figure}
\begin{center}
\begin{tabular}{c}
\includegraphics[width=0.6\textwidth]{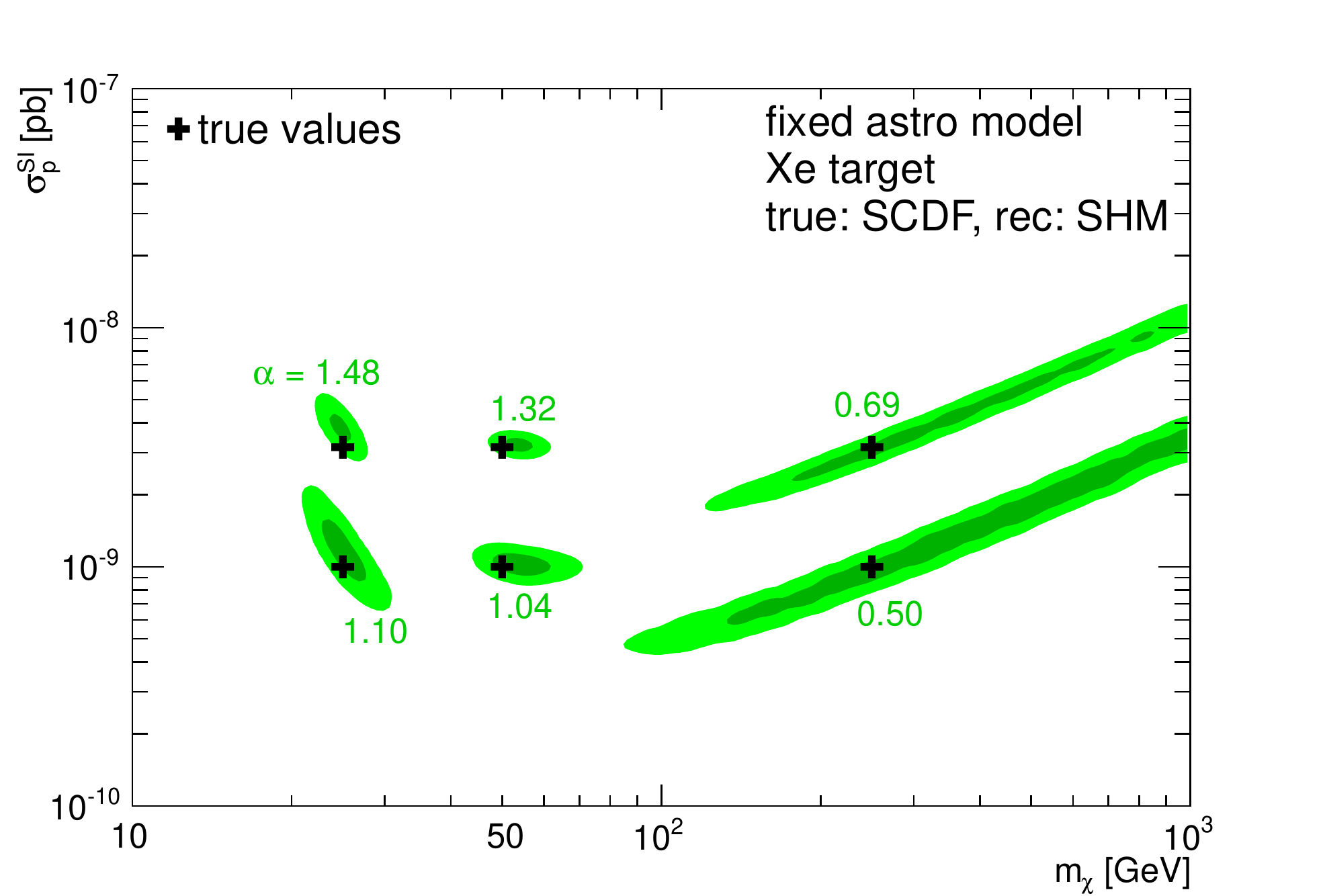}
\end{tabular}
\end{center}
\caption{Bias in the reconstruction of WIMP properties in a xenon ton-scale instrument for the fixed astrophysical setup, when the reconstruction assumes the standard halo model while the data are generated from the self-consistent distribution function with the fiducial parameters specified in Table \ref{tab:models}. Inner and outer contours correspond to the joint 68\% and 95\% posterior probability contours, respectively.  The parameter $\alpha$ quantifies the systematic bias according to Eq.~\eqref{bias}.
}
\label{fig:fixed}
\end{figure}

 We now study the systematic bias arising from assuming the wrong velocity distribution model in the reconstruction -- for reference, the integral in Eq.~\eqref{rate} (which controls the impact of the different velocity distribution models on the scattering rate) is shown in Fig.~\ref{fig:comparison} (right) in the case of the SHM and the SCDF. Fig.~\ref{fig:fixed} shows an example of the bias that can be incurred by assuming the wrong model for the velocity distribution. In this case, the underlying true event rate distribution is simulated from the SCDF, while the reconstruction is done using the SHM velocity distribution. The galactic parameters are fixed to their fiducial values in Table \ref{tab:models}. Fig.~\ref{fig:fixed} indicates a moderate bias in the reconstruction (typically around the $1\sigma$ level, as quantified by the bias parameter $\alpha$) for dark matter masses below 100 GeV.  This bias reduces significantly for WIMP masses larger than 100 GeV, but this is a consequence of the statistical errors becoming much larger at those mass values, as the mass--cross-section degeneracy means that only a lower bound can be placed on the WIMP mass. The relatively small bias in Fig.~\ref{fig:fixed} is partly due to the similar shape of the SHM and the SCDF (see left panel of Fig.~\ref{fig:comparison}). Note again that this statement is specific to the distribution function model that we consider; for distributions with multiple features the bias is more difficult to quantify at this time.

\begin{figure}
\begin{center}
\begin{tabular}{c}
\includegraphics[width=0.5\textwidth]{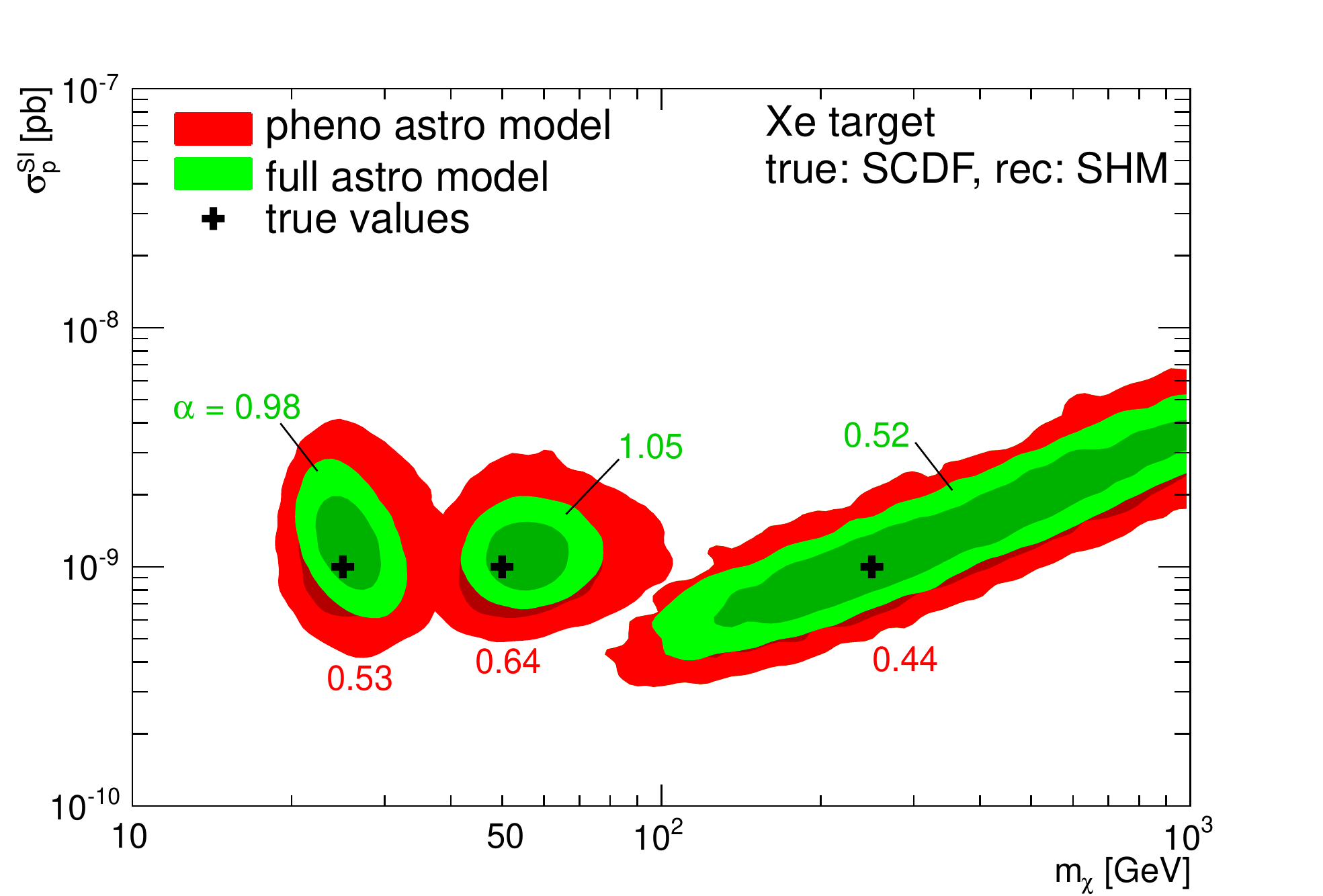}
\includegraphics[width=0.5\textwidth]{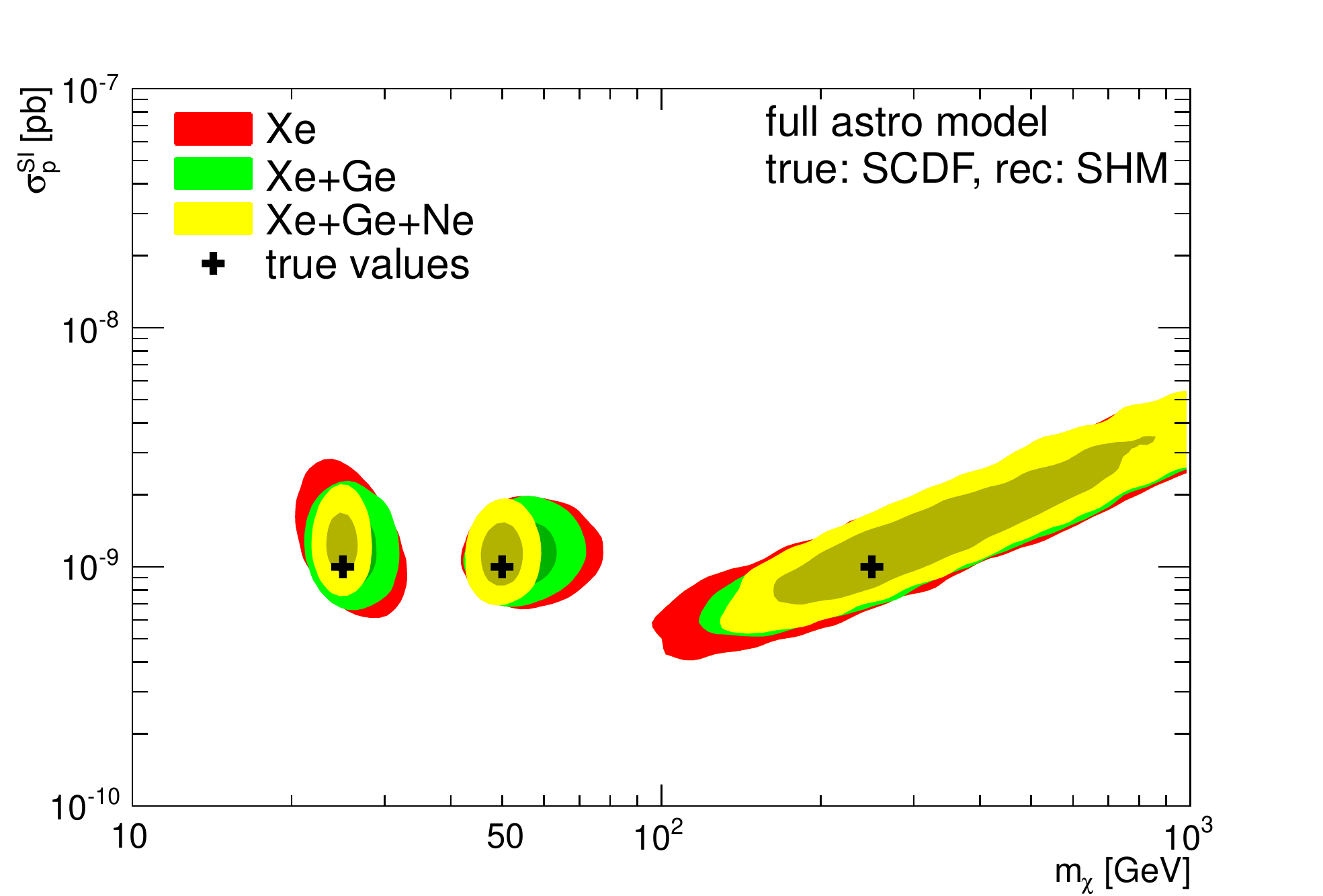}
\end{tabular}
\end{center}
\caption{Reconstruction of  WIMP properties when (wrongly) assuming the standard halo model while the data are generated from the self-consistent distribution function, but additionally marginalizing over astrophysical uncertainties. The left frame shows the case of a xenon ton-scale instrument for the pheno and full astrophysical setups, while the right frame corresponds to the combination of xenon, germanium and neon data for the full astrophysical setup. Inner and outer contours correspond to the joint 68\% and 95\% posterior probability contours, respectively. The systematic bias, as quantified by $\alpha$, is reduced with respect to Fig.~\ref{fig:fixed}, but statistical uncertainties increase. 
}
\label{fig:fullbase}
\end{figure}

\par Fig.~\ref{fig:fullbase} shows the reconstruction for three WIMP benchmarks, but expands on the analysis of Fig.~\ref{fig:fixed} by further marginalizing over galactic parameters. This is a more realistic scenario, where uncertainties in the astrophysical parameters are fully included in the inference about the WIMP properties by marginalizing over them.  Encouragingly, we find that there is no significant bias in the reconstruction in this case even when assuming the wrong velocity distribution model. Thus, for the models of the velocity distribution that we compare, assuming the wrong velocity distribution does not significantly bias the mass reconstruction. Of course, the price to pay for reducing the systematic bias is an increased statistical error on the WIMP parameters (compare with Fig.~\ref{fig:fixed}). Furthermore, we notice that marginalizing over astrophysical uncertainties in the more restricted pheno setup is sufficient to largely eliminate systematic bias, although the corresponding statistical uncertainty is higher than in the full setup. In comparing the left and the right panels of Fig.~\ref{fig:fullbase}, we see that the addition of different nuclear targets improves the reconstruction for all benchmarks.

\par
\begin{figure}
\begin{center}
\begin{tabular}{c}
\includegraphics[width=0.6\textwidth]{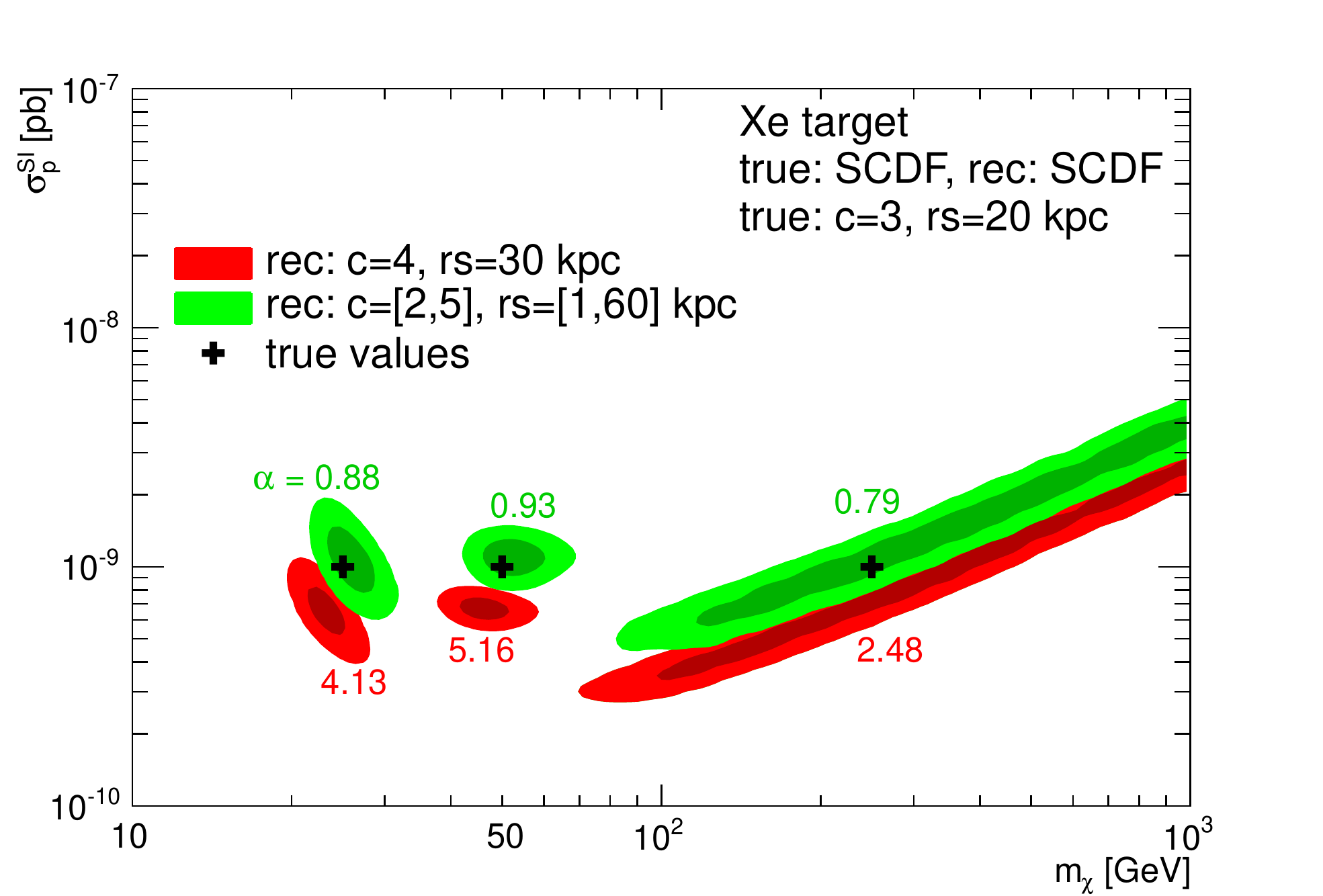}
\end{tabular}
\end{center}
\caption{Reconstruction of WIMP properties in a xenon ton-scale instrument when assuming a wrong outer slope $c$ and scale radius $r_s$. Red (dark) contours are for data generated using the self-consistent distribution function with $c=3$ and $r_s=20$ kpc, while the reconstruction wrongly assumes $c=4$ and $r_s=30$ kpc (all other galactic parameters are kept fixed). Green (light) contours result from the same procedure but additionally marginalizing over the parameters $c\in\left[2,5\right]$ and $r_s\in\left[1,60\right]$ kpc (with flat priors), thus significantly reducing systematic bias, as quantified by the parameter $\alpha$.
}
\label{fig:slopes}
\end{figure}

 In order to illustrate further the importance of accounting for astrophysical uncertainties for each of the models we consider, in Fig.~\ref{fig:slopes} we show the bias in the reconstruction arising from assuming the wrong value for the outer slope $c$ of the dark matter density profile and the scale radius $r_s$ when determining the velocity distribution through the Eddington formula. We have chosen the outer slope and the scale radius because those are two of the most significant parameters that impact the velocity distribution in our formalism \cite{Lisanti:2010qx,Mao:2012hf}. Though both halo models provide a good description to the galactic halo \cite{Lisanti:2010qx}, it is clear that they imply very different underlying values for the WIMP properties. As above, the reconstruction is seen to be more biased in this case for lower mass WIMPs. We do find, however, that marginalizing over the outer slope and scale radius does reduce this bias, as shown by the upper sets of green (light) contours in Fig.~\ref{fig:slopes}. This simple procedure reduces the systematic bias by up to a factor of 5, thus significantly improving the accuracy of the reconstruction.

\par
\begin{figure}
\begin{center}
\begin{tabular}{c}
\includegraphics[width=0.6\textwidth]{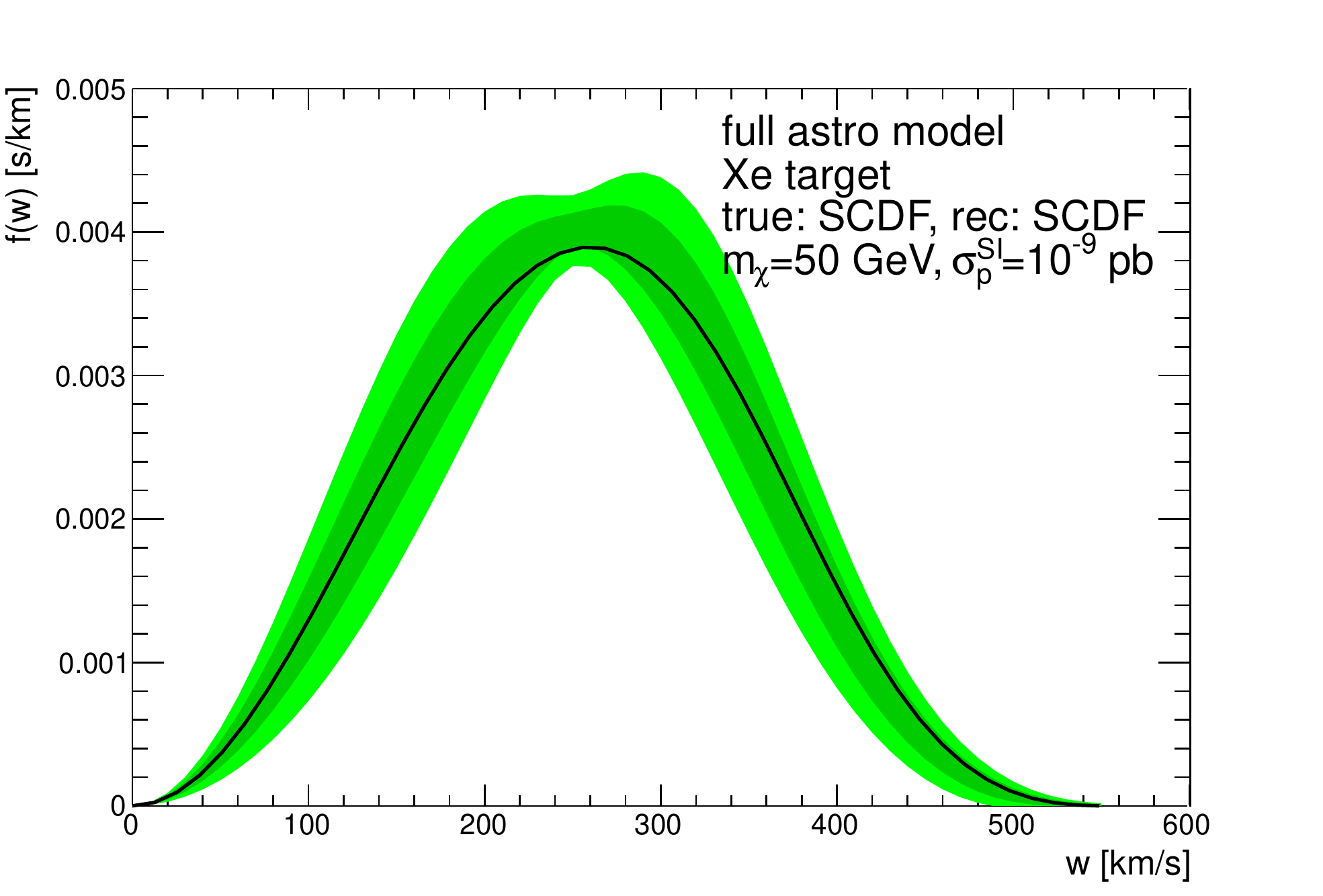}
\end{tabular}
\end{center}
\caption{Reconstruction of the velocity distribution function for a ton-scale xenon detector and a $50$ GeV WIMP. The inner and outer bands indicate the 1 and 2$\sigma$ ranges around the mean of the marginal posterior, respectively. The true velocity distribution is shown by the solid line.
}
\label{fig:recf}
\end{figure}

 In the context of our isotropic velocity distribution model, it is also possible for us to ask how well we can reconstruct the velocity distribution with an observed event rate distribution. In Fig.~\ref{fig:recf} we show the reconstructed velocity distribution assuming a $50$ GeV mass WIMP and a ton-scale xenon detector. Overall we find that the reconstruction is unbiased at all velocities, so it is possible to determine the velocity distribution from direct detection data for these parameters, at least in parametric form. We also found that the marginal posterior distribution at each velocity value is fairly close to Gaussian. Note that the measured event rate does not directly probe the velocity distribution below the minimum velocity to scatter at a given energy in xenon. Therefore, the parametric reconstruction presented here relies somewhat on the specific class of models we assumed. However, it is encouraging to note that, in this context, the reconstruction of the velocity distribution is unbiased.

\section{Conclusion}\label{secConc}

\par In this paper we have addressed the question of the bias in the determination of the WIMP mass and spin-independent cross-section for plausible models of the velocity distribution function. For the two specific models that we consider, the standard Maxwell-Boltzmann distribution and isotropic, spherically symmetric model determined from the Eddington formula, we have shown that assuming the wrong velocity distribution model only leads to moderate bias in the determination of WIMP parameters. This can be further mitigated by using astrophysical data to constrain galactic model parameters and properly marginalizing over them. In addition to the different parameterizations used for the velocity distribution function in our modeling, we use kinematic data to marginalize over halo model parameters in determining the velocity distribution. Thus this paper represents an important step towards connecting WIMP event rates to a full galactic halo model in a theoretically self-consistent framework. 

\par While the focus of our analysis has been on two particular models of the velocity distribution, our method can be applied to analytic models that depend on both the energy and the angular momentum \cite{Ullio:2000bf}. Though these models add degrees of freedom to the galactic model analysis, it is not clear whether they are a better description of our dark matter halo. However, anisotropic models do admit interesting theoretical phenomenology, allowing for potential model-independent extraction of the anisotropy of the WIMP velocity distribution \cite{Alves:2012ay}. 

\par This paper focused on the simplest elastic scattering dark matter models for detectors without directional sensitivity. In the future it will also be interesting to consider directional signals. Initial analysis along these lines have been undertaken assuming phenomenological models of the velocity distribution \cite{Lee:2012pf}. Connecting these distributions to both galactic halo model parameters and to the dark matter distribution in the Galaxy, as we have done here for isotropic models, will likely provide more rigorous constraints on the three-dimensional WIMP distribution \cite{Morgan:2004ys}. Further, it will be interesting to apply our methods to annual modulation signals \cite{Lewis:2003bv}. Also, in the future it will be important to apply the method that we have outlined to more general parameterizations of the elastic scattering cross-section and spin-dependent cross-sections \cite{Chang:2009yt,Fan:2010gt}. In fact, spin-dependent couplings induce different dependencies of the event rate on the nuclear recoil energy and can in many cases lead to stronger bounds than spin-independent scattering \cite{Garny:2012it}.

A general conclusion of our study is that the use of the standard Maxwellian velocity distribution leads to a robust determination of the WIMP mass from direct detection experiments, as long as astrophysical uncertainties are kept into account through marginalization over galactic model parameters.

\acknowledgments 
We thank Yao-Yuan Mao and Risa Wechsler for useful discussions. M.P.~is indebted to the Institute for Theoretical Physics at the University of Zurich and acknowledges the support from the Swiss National Science Foundation in the early phases of this work. M.P.~would also like to thank the hospitality of the Astrophysics Group at Imperial College London. R.T.~would like to thank the Institute for Theoretical Physics at the University of Zurich and the GRAPPA Institute at the University of Amsterdam for hospitality. G.B.~acknowledges the support of the European Research Council through the ERC Starting Grant {\it WIMPs Kairos}.


\bibliographystyle{JHEP}
\bibliography{astrobias}

\end{document}